\documentclass[%
 aip,
 amsmath,amssymb, longbibliography,
preprint,%
 reprint,%
]{revtex4-1}

\usepackage[breaklinks]{hyperref}
\usepackage{graphicx,xcolor}
\usepackage{dcolumn}
\usepackage{bm}

\usepackage[utf8]{inputenc}
\usepackage[T1]{fontenc}
\usepackage{mathptmx}

\newcommand{\vecform}{\bm}

\newcommand{\nn}{\vecform{n}}
\newcommand{\bra}[1]{\mbox{$\langle #1 |$}}

\newcommand{\ket}[1]{\mbox{$| #1 \rangle$}}

\newcommand{\rr}{\bm{r}}

\begin{document}

\preprint{AIP/123-QED}

\title{On the time evolution of fermionic occupation numbers}

\author{Carlos L. Benavides-Riveros}
\email{carlos.benavides-riveros@physik.uni-halle.de}
\author{Miguel A. L. Marques}%
 \affiliation{Institut f\"ur Physik, Martin-Luther-Universit\"at
Halle-Wittenberg, 06120 Halle (Saale), Germany}

\date{\today}

\begin{abstract}
We derive an equation for the time evolution of the natural 
occupation numbers for fermionic systems with more than two 
electrons. The evolution of such numbers is connected with the 
symmetry-adapted generalized Pauli exclusion principle, as 
well as with the evolution of the natural orbitals and a set 
of many-body relative phases. We then relate the evolution of 
these phases to a geometrical and a dynamical term, attached to 
each one of the Slater determinants appearing in the 
configuration-interaction expansion of the wave function.
\end{abstract}

\maketitle

\section{Introduction}

The time evolution of an electronic system is governed by 
the Schr\"odinger equation~\cite{PhysRev.28.1049}. Yet 
a real-time propagation of the many-fermion wave function 
is, by and large, computationally prohibitive.  The 
time-dependent extension of density functional theory 
(TDDFT) alleviates this computational problem by mapping 
the evolution of the ground-state density 
to the one of a certain auxiliary system \cite{RungeGross}. 
Since such auxiliary system is \textit{non-interacting}, 
TDDFT does not involve fractional occupation numbers, 
which are at any rate important 
for capturing quantum correlations~\cite{appel, bresinova}, 
even in the adiabatic regime \cite{requist}.

It is well known that the ground-state wave function of an 
electronic system can be written as a functional of the 
one-body reduced density matrix, which for a wave function 
$\Psi$ is defined as
\begin{align}
\gamma(1,1') \equiv \bra{\Psi}\hat \psi^\dagger(1')
\hat \psi(1)\ket{\Psi},
\end{align}
with the short notation $1\equiv (\rr_1,\varsigma_1)$ for 
position and spin coordinates. Thus, ground states can be 
viewed as functionals of such reduced densities (say, 
$\Psi_{\rm gs}[\gamma]$)
\cite{PhysRevB.12.2111}. 
Since $\gamma$ accounts for fractional na\-tu\-ral occupation 
numbers (i.e.,~its eigenvalues), employing this matrix as 
the main object leads to a theory able to capture quite 
well strong (static) electron correlations 
\cite{Cohen792,Sanchez,C7CP01137G,Schade2017}. For instance, 
unlike density functional theory, such a density-matrix functional 
theory correctly predicts the insulating 
behavior of Mott-type insulators \cite{Sharma, Pernalrev}.
Furthermore, it has been recently pointed out that $\gamma$ 
encodes essential many-body aspects of interacting fermions 
and hard-core bosons, as many-body localization transitions 
\cite{M1, M2}, entanglement \cite{Walter1205, Maci_ek_2018},
or topological 
states \cite{M3}. Given these remarkable properties, there 
is a growing interest in proposing protocols to access 
to the structure of this density matrix both experimentally 
using quantum-gas microscopes \cite{M4}, or theoretically 
employing quantum computers \cite{MazziottiQC} and hard-core 
bosons \cite{CSHCB}.

Unfortunately, time-de\-pe\-ndent extensions of the theory of 
the one-body reduced density matrix suffer from various 
shortcomings. Save for two-electron systems, the current 
status of the theory does not allow the fer\-mionic 
occupation numbers to evolve in time 
\cite{Pernal, GBG08,vanmeer,Klaas}. 
To understand the problem it is worth recalling that the 
Schr\"odinger equation leads to the BBGKY hierarchy, 
whose equation for $\gamma(t)$ is \cite{Bonitz}:
\begin{align}
\label{BBGKY}
&i  \frac{d\hat \gamma(t)}{dt} = 
[\hat h(t), \hat \gamma(t)] + \hat u(t),
\end{align}
where $\hat h(t)$ is the time-dependent one-particle
Hamilto\-nian operator and $\hat u(t)$ is in spatial and
spin represen\-ta\-tion
$u(1,1',t) =
2 \int [v(1,2) - v(1',2)] \Gamma(1,2;1',2,t) d2$.
$v(1,2)$ is the Coulomb potential and 
$\Gamma(1,2;1',2',t)$
is the two-particle reduced density matrix. 
The density matrix $\hat\gamma(t)$ could 
also be com\-puted by integrating out $\Gamma(t)$
which satisfies in turn an equation similar to~\eqref{BBGKY}. 
Since the representability conditions of the 
three-particle reduced density matrix $\Gamma_3$ 
are much harder to implement,
this latter procedure is not completely well-defined:
the positive-semidefiniteness of $\Gamma_3$
is not necessarily inherited by $\Gamma$ and 
$\gamma$, neither the energy 
is conserved in the absence of time-dependent 
potentials \cite{PhysRevB.85.235121}. For this 
reason, it is believed that fermionic constraints on the 
occupation numbers should play a role in implementing 
the BBGKY hierarchy \cite{bresinova}. 

By definition, 
in the natural-orbital basis $\hat \gamma (t)$ is diagonal, reading
\begin{align}
\hat \gamma(t) = \sum_k n_k(t) \ket{\varphi_k(t)}\bra{\varphi_k(t)}.
\end{align}
By multiplying Eq.~\eqref{BBGKY} by $\varphi^*_j(1)$ and 
$\varphi_k(1')$, and integrating both position 
and spin coordinates, Pernal, Gritsenko and Baerends
obtained an equation for the time-evo\-lution for the natural orbitals \cite{Pernal}, 
namely,
\begin{align}
\label{eq:NONt}
i\bra{\varphi_j} \dot \varphi_k\rangle =   
\bra{\varphi_j} \hat h \ket{\varphi_k}  
 +\frac{W_{jk} - W_{kj}^*}{n_k - n_j}, \qquad 
 j\neq k
\end{align}
as well as an equation for the time evolution 
of the natural occupation numbers:
\begin{align}
\label{eq:non}
\dot n_k =   2 \Im [W_{kk}],
\end{align}
where $W_{jk} \equiv 2 \int  
v(1,2) \Gamma(1,2;1',2) \varphi^*_j(1) \varphi_k(1') d1 d1' d2$. 
The presence of the difference $(n_k - n_j)$ in Eq.~\eqref{eq:NONt}
indicates that degeneracies of the occupations lead to 
singularities in the time evolution of the natural orbitals. 
It is not a surprise, since degenerancy of the occupation 
numbers implies an ambiguity in the definition of the 
corresponding natural orbitals (for a linear combination of 
degenerate natural orbitals is also a natural orbital). 
For simplicity, it is in general assumed the absence of 
such a degenerancy.
Since the imaginary part of $W_{kk}$ determines the 
time evolution of the occupations \eqref{eq:non}, it is clear that 
some relative phases are crucial to correctly capture 
the dynamics of the system. With the exception of the 
L\"owdin-Shull functional for two-ele\-ctron 
systems~\cite{LS}, the PNOF4 functional for density-matrix 
functional theory~\cite{doi:10.1002/qua.24020}, and its latter 
developments~\cite{Mitxelena2018}, the right-hand 
of Eq.~\eqref{eq:non} vanishes identically for current 
reconstructions of $\Gamma$ in terms of $\gamma$, so the 
occupation numbers do not evolve in 
time \cite{Pernal2016}. There are a few attempts in the 
literature to account for relative phases at the 
level of the two-body reduced density matrix. Yet, 
the theory developed in this way is limited to the 
two-electron case 
\cite{GGB10, doi:10.1063/1.3687344, requisttwo, PhysRevA.90.012518}.
By studying the underlying exchange symmetry, 
this paper is aimed at proposing a new way 
of tackling the time dependency of the natural 
occupation numbers of fermionic systems. By doing so,
we present an approximative formula for the adiabatic time
evolution of the occupation numbers for fermionic 
systems. 

Besides this introduction, the paper contains three 
additional parts. In Sec.~\ref{sec2} we discuss the 
so-called generalized Pauli constraints and how they 
can help us to extract information of the wave function. In Sec.~\ref{sec3}, 
we present a couple of formulas for the time evolution of 
a pinned system 
of three electrons in six natural orbitals. 
By exploiting the information of symmetries, Sec.~\ref{sec4}
generalizes this result for larger systems. The paper ends with 
a conclusion section and three appendices.

\section{Robustness of fermionic constraints}
\label{sec2}

\begin{figure}
\includegraphics[width=5cm]{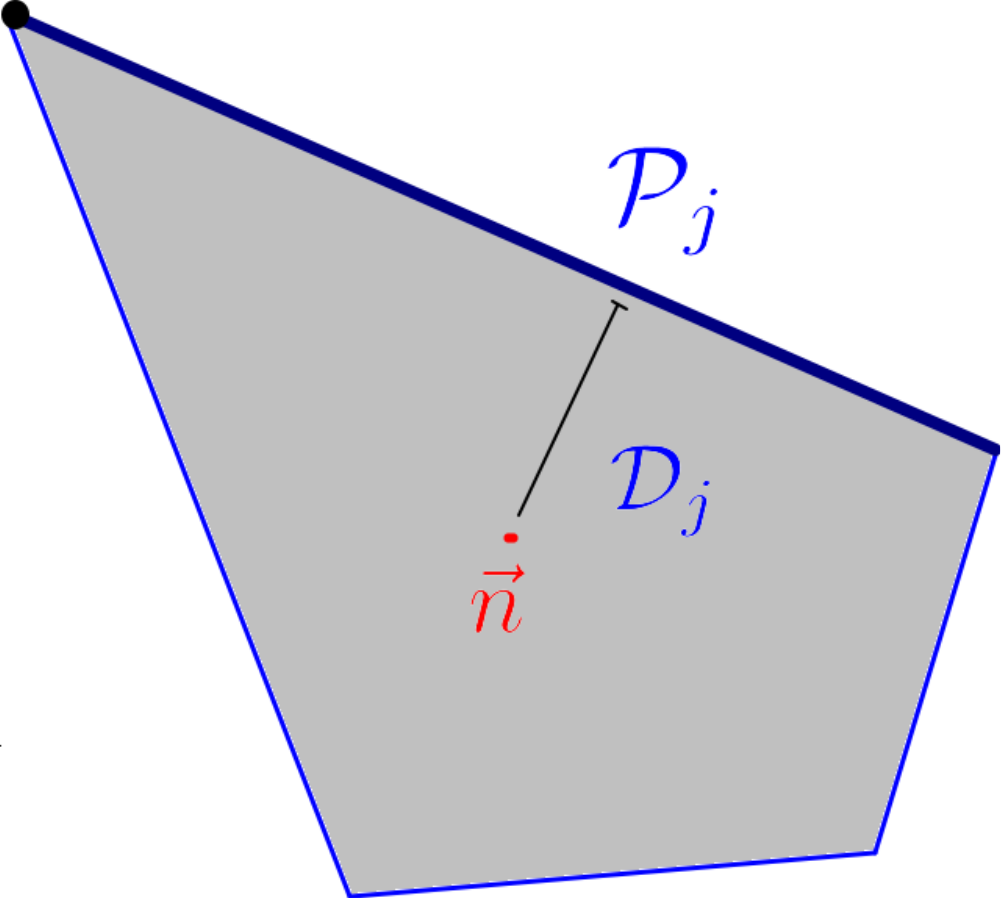}
  \caption{Illustration of the distance $\mathcal{D}_j$ of $\nn$ 
  to  the  polytope  facet $\mathcal{P}_j$. }
  \label{figure}
\end{figure}

To first shed light into this problem we study the evolution 
of the one-body reduced density matrix for pinned wave 
functions, which, as we will show, are structural
simplifications of some ground states. From a 
formal viewpoint, it is known that the compatibility, 
or represen\-tability, of a fermionic one-body reduced 
density matrix $\gamma$ with respect to a quantum 
many-body state $\ket{\Psi}$ is described by sets of 
linear constraints on its spectra 
$\nn \equiv (n_1, n_2, \dots)$, namely  \cite{Kly2,Kly3}:
\begin{equation}
  \label{eq:gpc}
  \mathcal{D}_j(\nn) \equiv \kappa_j^{0}+\sum_{i}\kappa_j^{i}  
  n_i\geq 0 ,
\end{equation}
where the coefficients $\kappa_j^{i}$ are integers depending on the 
number of fermions $N$ and the dimension of the underlying one-particle 
Hilbert space $M$. Along with the non-increasing ordering of the natural 
occupation numbers 
(say, $n_i \geq n_{i+1}$) and the sum rule ($\sum_i n_i = N$), 
these \textit{generalized Pauli constraints} \eqref{eq:gpc} 
define a polytope where the sets of $\nn$, which are compatible 
with $N$-fermion pure states, lie \cite{doi:10.1142/S0129055X14500044}. 
It is quite remar\-ka\-ble that,
whenever so\-me of those quantum marginal constraints are saturated 
or \textit{pinned}, the total quantum state has a specific, simplified 
structure. Indeed,
$\mathcal{D}_j(\nn)  = 0   \Longleftrightarrow  \hat D_j \ket{\Psi} = 0$,
where the operator $\hat D_j =
\kappa_j^{0}+\sum_{i}\kappa_j^{i}  
\hat n_i$ is built by replacing the occupation numbers in 
Eq.~\eqref{eq:gpc} by the corresponding particle number 
operators. The importance of this result lies on the 
fact that it provides an important selection rule for the Slater 
determinants that can appear in the configuration-interaction expansion 
of wave functions \cite{CSHFZPC}. Indeed, a wave function 
whose spectrum is pinned to one of the polytope's facets 
$\mathcal{P}_j = \{\nn| \mathcal{D}_j(\nn) = 0\}$ (see Fig.~\ref{figure})
can be written as a linear superposition of the Slater 
determinants which belong to the zero-eigenspace of the 
operator $\hat D_j$. This selection rule can be used to 
systematically produce ans\"atzen for ground states in the 
form of sparse wave functions, which, instead of using the 
full Hilbert space, can be expanded in the basis of the natural 
orbitals (the eigenvalues of $\gamma$) with a few Slater 
determinants \cite{CSHFZPC,Chakra}. Apart from the 
simplification of the wave function, there is another 
advantage in using natural orbitals that is worth 
mentioning here: it is known that the basis of natural orbitals 
is typically quite good to convergence the full wave function.

This structural simplification for pinned 
quantum systems is stable in the sense that  
any many-fermion quantum state can be 
approximated by the structural simplified form corresponding 
to saturation of the generalized Pauli constraint 
$\mathcal{D}_j$. The error of such a simplification is 
boun\-ded by twice the distance of the vector of its 
occupation numbers to the corresponding 
polytope's facet $\mathcal{P}_j$ \cite{SBV}. 
Recently, it has been suggested that the generalized Pauli 
constraints may facilitate the development of more 
accurate functionals within density-matrix 
functional theory \cite{Halle, doi:10.1063/1.5080088, 
PhysRevLett.122.013001}. Since quasipinning 
(say, $\mathcal{D}_j(\nn) \approx 0$) is approximately 
observed for several ground states, the quasipinning 
``mechanism'' has attrac\-ted some attention in quantum 
chemistry and quantum-information theory \cite{CSQMath12, 
progress, CS2013, BenavLiQuasi, CSQuasipinning, Mazz14, 
MazzOpen, BenavQuasi2,RDMFT,recentMazziotti, TVS16, CS2016b,DePrince,
TVS17, newpaper, PhysRevA.97.052503,doi:10.1021/acs.jpclett.8b03028}.  
 
It can also be shown that the structural simplification of 
the wave function is also stable in the sense that pinning is
robust under any small perturbation of the Hamiltonian. 
Klyachko has indeed suggested that a pinned system should remain 
so under a reasonably small variation of the Hamiltonian \cite{Kly1}.
This can be easily seen by perturbing a Hamiltonian $\hat H$ with 
non-degenerated eigenstates $\ket{\Psi_n}$ and eigenenergies $E_n$. 
In perturbation theo\-ry the ground state 
of the perturbed Hamiltonian $\hat H(\lambda) \equiv \hat H 
+ \lambda \hat V$ reads as $\ket{\Psi^{\lambda}_0} = \ket{\Psi_0} 
 - \lambda \sum_{n\neq 0} b_n 
\ket{\Psi_n} + \mathcal{O}(\lambda^2)$, with 
$b_n = \bra{\Psi_n}\hat V\ket{\Psi_0}/(E_n - E_0)$.
If the unperturbed ground state is pinned to a facet $\mathcal{P}_j$, 
then $\hat D_j \ket{\Psi_0} = 0$, and therefore, the perturbed distance to
that polytope's facet reads now as
${\mathcal D}_j(n^\lambda_0) = \bra{\Psi^\lambda_0}\hat D_j\ket{\Psi^\lambda_0} 
\sim \mathcal{O} (\lambda^2)$.
Based on a self-consistent perturbation theory, it has also been shown 
that a perturbation of a one-particle Hamiltonian (whose ground state is 
a Slater determinant) induces a change in pinning only in second order 
\cite{PhysRevA.97.052105}. Those  results are somehow expected, since
the expectation value of any symmetry $\hat{\mathfrak{s}}$ satisfied by 
a (non-degenerated) ground state remains constant in first-order pertubation theory. Indeed, 
\begin{align}
\bra{\Psi^\lambda_0}\hat{\mathfrak{s}}\ket{\Psi^\lambda_0} 
= \bra{\Psi_0}\hat{\mathfrak{s}}\ket{\Psi_0} + \mathcal{O} (\lambda^2).
\end{align}
This immediately implies that whenever the (non-dege\-ne\-ra\-ted) ground state 
of a system belongs eigenspace of a given symmetry operator, it remains in such an 
eigenspace in first order upon perturbation. For instance, as is well known, 
a system which is 
pinned to the Pauli exclusion principle (say, a occupancy is equal to 1 or 0) 
stay so in the first order of the perturbation.

Remarkably the same is \textit{approximately} true for quasipinning. In Appendix 
\ref{appaa} we show that the response of quasipinning under a perturbation 
is bounded from above by the formula:
\begin{align}
\label{eq:newk}
\mathcal{D}_j(\nn^\lambda_0) \leq \mathcal{D}_j(\nn_0)
 + 2 \lambda \gamma_V \sqrt{\langle\hat D^2_j \rangle}
+ \mathcal{O} (\lambda^2),
\end{align} 
where $\langle\hat D^2_j \rangle = 
\bra{\Psi_0}\hat D^2_j\ket{\Psi_0}$. The multiplicative 
prefactor $\gamma_V$ is 
a relative strength between the perturbation and the 
unpertur\-bed Hamiltonian. This strength can be further 
bounded (see Appendix \ref{appaa}):
\begin{align}
\gamma_V \leq \frac{\sqrt{{\rm Cov}_{\Psi_0}(\hat V)}}{E_{\rm gap}} ,
\end{align}
where the energy gap is defined as the difference between
the first-excited and ground-state energies 
$E_{\rm gap} = E_1 - E_0$,
and the covariance of the perturbation is
${\rm Cov}_{\Psi_0}(\hat V) =
\bra{\Psi_0}\hat V^2 \ket{\Psi_0}
- \bra{\Psi_0}\hat V \ket{\Psi_0}^2$. 

The appearance of
the term
$\langle\hat D_j^2 \rangle$ in \eqref{eq:newk} indicates
that, in or\-der to predict the stability of quasipinned 
systems, the expected value of the square of the operator 
$\hat D_j$ deserves further attention in quasipinning theory.
It is worth noticing that, when the system is pinned, 
$\langle\hat D^2_j \rangle = 0$, and therefore  
the pinning response goes only in second order, as Klyachko
correctly stressed. In a more general fashion, 
paraphrasing Klyachko, we have shown that since a quasipinned system 
is \textit{approximately} driven by Pauli kinematics, it should remain 
\textit{approximately} pinned to that facet under a reasonably small variation 
of the Hamiltonian, as long as $\langle\hat D_j^2 \rangle$ is also small.

\section{A formula for the time evolution of the fermionic occupation numbers}
\label{sec3}

This robustness of (quasi)pinning prompts us to seek for
the equation of motion of the one-body reduced 
density matrix of a pinned quantum system. Geometrically, 
the aim is to constrain the dynamics of the system to 
move on a hyperplane in the one-particle picture. 
To illustrate our approach let us consider the so-called 
Borland-Dennis wave function, namely, the pinned rank-six 
approximation (i.e., $M = 6$) 
for the three-active-electron system \cite{Borland1972}. 
This wave function is known to be pinned to one of the 
facets of the corresponding polytope 
(i.e., $n_1 + n_2 + n_4 = 2$) and can be 
written explicitly in terms of the amplitude squares $f_\alpha$, 
the natural orbitals $\varphi_\alpha$ (whose time derivative is defined by 
Eq.~\eqref{eq:NONt}) and the relative phases $\xi_\alpha$. It reads:
\begin{align}
\ket{\Phi_{\rm BD}[\vecform{f},\vecform{\varphi},\vecform{\xi}]} =   
\sum_\alpha \sqrt{f_\alpha} 
e^{-i\xi_\alpha}\ket{\vecform{\varphi_\alpha}}, 
\label{BD}
\end{align}
where $\ket{\cdot}$ denotes normalized Slater determinants. The 
coefficient
 $\alpha \in \{3,5,6\}$, and 
$\vecform{\varphi_3}(t) \equiv 
\varphi_1(t)\varphi_2(t)\varphi_3(t)$, $\vecform{\varphi_5}(t) 
\equiv \varphi_1(t)\varphi_4(t)\varphi_5(t)$ and 
$\vecform{\varphi_6}(t) 
\equiv \varphi_2(t)\varphi_4(t)\varphi_6(t)$. 
In the Borland-Dennis state \eqref{BD} the square of the 
amplitudes $\vecform{f}$
turn out to be equal to three occupation numbers. This trivial
correspondence between $\nn$ and $\vecform{f}$ will be crucial 
to later generalize our results. For the wave function \eqref{BD} there are other three 
saturated generalized Pauli constraints, namely, $n_k + n_{7-k} = 1$ 
with $k \in \{1,2,3\}$.  
We will exploit later the well-known fact that the relative
phases $\xi_\alpha$ 
can only be uniquely defined with respect to a given choice of the 
time-dependent phases of the natural orbitals \cite{requisttwo}.

Recall that the time evolution of the natural orbitals is 
completely determined by the off-diagonal terms 
$\bra{\varphi_j} \partial_t \varphi_k\rangle$ $(j \neq k)$
in Eq.~\eqref{eq:NONt}. The expression for the two-particle 
reduced density matrix can be easily found by tracing out 
1 particle from the full density matrix:
$\Gamma = 3 {\rm Tr}_{1} [\ket{\Phi_{\rm BD}}\bra{\Phi_{\rm BD}}]$.
The missed diagonal terms
$\bra{\varphi_k} \dot \varphi_k\rangle$ can be removed as 
convenient phase factors 
\cite{doi:10.1063/1.4974096} (see below). 
The equations of motion of $\vecform{\xi}$ and $\vecform{f}$
can be de\-rived from the stationary condition of 
the time-dependent quan\-tum mechanical
action 
\begin{align}
\mathcal{A}_\Psi(t_1,t_2) = 
 \int_{t_1}^{t_2} \bra{\Psi(\tau)} i\partial_\tau - \hat H(\tau) 
 \ket{\Psi(\tau)} d\tau. 
\end{align}
That this action gives place to the Schrödinger operator
can be easily seen by variating $\mathcal{A}_\Psi$ 
with respect to the state $\bra{\Psi(t)}$ under the constraint of 
norm conservation \cite{LOWDIN19721}. $\mathcal{A}_\Psi$ is 
therefore stationary for the correct state $\ket{\Psi(t)}$ which 
develops from a given initial state 
$\ket{\Psi(t_1)}$.
By optimizing the functional $\mathcal{A}_{\Phi_{\rm BD}}$
with respect to the phases, one obtains an explicit equation for 
the evolution of the square amplitudes. 

The time derivative of the Borland-Dennis 
wave function \eqref{BD} gives
$\bra{\Phi_{\rm BD}}i \partial_t\ket{\Phi_{\rm BD}} = 
  \sum_\alpha f_\alpha (\dot\xi_\alpha  + \bra{\vecform{\varphi_\alpha}}
i\partial_t \ket{\vecform{\varphi_\alpha}})
$.
We have used the fact that $\sum_\alpha \dot f_\alpha = 0$ and 
$\bra{\vecform{\varphi_\beta}}\partial_t\ket{\vecform{\varphi_\alpha}} = 0$
for $\beta \neq \alpha$, 
because two different Slater determinants in 
Eq.~\eqref{BD} differ by at least
two orbitals. The expected value of the Ha\-mil\-tonian is:
\begin{align}
\bra{\Phi_{\rm BD}} \hat H \ket{\Phi_{\rm BD}} =
\sum_{\alpha\beta} \sqrt{f_\alpha f_\beta} e^{i(\xi_\beta - \xi_\alpha)}
\bra{\vecform{\varphi_\beta}} \hat H \ket{\vecform{\varphi_\alpha}}.
\end{align}
Optimizing the functional $\mathcal{A}[\Phi_{\rm BD}]$ 
with respect to the phases gives the following equation of motion:
\begin{align}
\label{eq:tdNON1}
i\dot{f}_\alpha = \sum_{\beta \neq \alpha} \sqrt{f_\alpha f_\beta}\left[
\bra{\vecform{\varphi_\beta}} \hat H \ket{\vecform{\varphi_\alpha}}
e^{i(\xi_\beta - \xi_\alpha)} - {\rm c.c.}\right].
\end{align}  
 To complete the time-evolution picture of the 
quan\-tum system we need equations for the evolution 
of the relative phases $\vecform{\xi}$, which 
can be derived from the stationary condition of 
the action $\mathcal{A}[\Phi_{\rm BD}]$ 
with respect to $\vecform{f}$. 
The evolution of the relative phases is determined by 
the instructive relation:
\begin{align}
\label{phases}
\xi_\alpha(t) =
\xi_\alpha^{\rm geo}(t)
+ \xi^{\rm dyn}_\alpha(t).
\end{align}
This result relates the evolution of the 
phases with a \textit{Slater-geometrical} phase 
\begin{align}
\label{eq:geo}
\xi_\alpha^{\rm geo}(t) = - i\int_0^t \langle\vecform{\varphi_\alpha}| 
\partial_\tau\ket{\vecform{\varphi_\alpha}}d\tau
\end{align} attached to the 
Slater deter\-minant $\ket{\vecform{\varphi_\alpha}}$, and an 
additional \textit{Slater-dynamical} phase which is written in terms 
of the diagonal and non-diagonal elements of the Hamiltonian driving
the dynamics of the system, namely,
\begin{align}
\xi^{\rm dyn}_\alpha(t) = \int_0^t \sum_{\beta} 
\frac12 \sqrt{\frac{f_\beta}{f_\alpha}} \left[
\bra{\vecform{\varphi_\alpha}} \hat H \ket{\vecform{\varphi_\beta}}
e^{i(\xi_\alpha - \xi_\beta)} + {\rm c.c.} \right] d\tau.
\label{coorph}
\end{align}
Notice that the Slater-geometrical phase 
$\xi_\alpha^{\rm geo}(t)$ contains the missing diagonal terms of Eq.~\eqref{eq:NONt}. 
This phase indicates clearly that the natural orbitals should be shifted 
accordingly.  To see that let us define the phase-shifted
natural orbitals
\begin{align}
\ket{\tilde{\varphi}_k} = e^{-\int_0^t \bra{\varphi_k} \partial_{\tau} \ket{\varphi_k} d\tau} \ket{\varphi_k}.
\end{align}
By construction (see Appendix \ref{appc}),
we have 
$\bra{\tilde\varphi_k}\tilde\varphi_l\rangle = \delta_k^l$ and in particular
$\bra{\tilde\varphi_k}\partial_t\ket{\tilde\varphi_k} = 0$,
which means that the derivative $\partial_t\ket{\tilde\varphi_k}$
is perpendicular to $\ket{\tilde\varphi_k}$. This is the 
parallel-transport well-known condition. In turn, 
Slater determinants satisfy 
\begin{align}
\label{eq:mirac}
\ket{\vecform{\varphi}_{\alpha}} = 
e^{i\xi^{\rm geo}_\alpha} \ket{\vecform{\tilde\varphi}_{\alpha}},
\end{align}
which is a parallel-transport condition for Slater determinants.
Notice the different sign in front of the phases in  
Eq.~\eqref{eq:mirac} and Eq.~\eqref{BD}. Therefore, 
in the basis of phase-shifted natural orbitals, the
Slater-geometrical phase does not contribute to the time 
evolution of the natural occupation numbers. We can now 
rewrite our results in terms 
of $\ket{\vecform{\tilde\varphi}_{\alpha}}$: 
formulas \eqref{eq:tdNON1} and \eqref{coorph} change just by 
replacing $ \ket{\vecform{\varphi}_{\alpha}} \rightarrow 
\ket{\vecform{\tilde\varphi}_{\alpha}}$
and $\xi_\alpha(t) \rightarrow \xi^{\rm dyn}_\alpha(t)$.
Since the seminal paper of Berry \cite{Berry}, this kind of 
pha\-ses has been discovered in many fields of physics \cite{PhysRevLett.58.1593,
  PhysRevLett.80.1800,RevModPhys.82.1959, PhysRevA.93.042108}. Yet the result
\eqref{coorph} is unique in establishing a direct 
relationship for fermionic systems between Slater determinants
and the dynamical phases. Moreover, since the wave function and 
the Slater determinants are written in terms of the 
natural orbitals and the occupation numbers, the 
phases presented here are functions of
one-body reduced quantities
(say, $\xi_\alpha^{\rm dyn}[\gamma,\vecform{\xi}^{\rm dyn}]$).  
We emphasize that Eq.~\eqref{coorph} is absent from the standard 
formulation of time-dependent density-matrix functional theory, 
and this is the reason for its severe shortcomings.
Notice also that the result \eqref{coorph} can be easily 
generalized to any natural orbital that appears in one 
and only one of the Slater determinants in the 
configuration-interaction expansion of the wave 
function (this is the argument for the two-electron 
system \cite{doi:10.1063/1.3687344}). In addition, it is 
remarkable that
the relative dynamical phases retain the memory effects of the system's
time evolution.

Before finishing this section, it is worth recalling 
that some ground states are very close to, 
but not exactly on, one of the boundaries of the polytope. 
For quasipinned systems we use as an ansatz a pinned wave 
function. Therefore,
by restricting the evolution to one of the hyperplanes 
in the one-particle picture, we are able to unveil 
an approximate equation for the evolution of the 
occupation numbers for a three-electron system.

\section{Generalization}
\label{sec4}

To generalize our results let us consider translationally invariant 
systems on a one-band lattice with periodic boundary conditions.
Let us consider, for instance, the Hubbard model with 
periodic boundary conditions. The magnetization $m$ and the total 
Bloch number $T$ are good quantum numbers. The Hamiltonian 
is block diagonal with respect to those symmetries (and other ones like
the total spin or the parity).
Clearly, the representability conditions for each symmetry sector are  
more  restrictive than the generalized Pauli exclusion principle, but  
the computation of the former constraints is considerably simpler 
than the calculation of the latter ones. This is the symmetry-adapted 
generalized Pauli exclusion principle, which we now exploit. To distinguish both types 
of generalized Pauli constraints, let us call $d^{s}_j(\nn)$ the ones 
coming from a symmetry-adapted sector $s$.

Due to translational and spin symmetries, 
any two Slater determinants belonging to the same symmetry sector
should differ by at least two natural orbitals \cite{PhysRevLett.122.013001}. 
It is a consequence of these symmetries that the corresponding 
one-body reduced density matrix is diagonal.
This kind of wave functions are used in quantum-computing 
simulations of quantum chemistry models \cite{PhysRevA.91.022311}.
Another important example of these wave functions is the seniority-zero 
sector of the Hilbert space for even-number of 
electrons \cite{doi:10.1063/1.3613706}. It has also been shown that writing 
the wavefunction in the basis of natural orbitals leads to a sharp drop 
of the coefficients of Slater determinants containing just single or triple excitations \cite{Mentel2014} which can also be 
argued by using pinning arguments for general systems \cite{BenavQuasi2}.

The wave function reads exactly
as \eqref{BD} but $\alpha$ stands now for a string of numbers indexing 
the natural orbitals in lexicographic order (e.g, 
$\ket{\vecform{\varphi}_{123}} = 
\ket{\varphi_1\varphi_2\varphi_3}$). 
The amplitudes $f_\alpha$ can be related with the natural occupation 
numbers by a linear transformation, $\mathcal{M}^s_{(N,M)}
 \vecform{f} = \nn$. The crucial
observation is that, whenever there are as many Slater determinants 
as independent occupation numbers,  
the matrix $\mathcal{M}^s_{(N,M)}$ is invertible \cite{PhysRevLett.122.013001}. 
Its inverse is a matrix of integers (up to a global normalization 
constant) whose entries depend on $(N,M)$ and the corresponding
symmetry sector. To give an example, consider the  Hubbard  model  with three  
spin-$\tfrac12$ fermions on four lattice  sites. For such a system, 
there are 31 $N$-representability conditions. Yet, after restricting to some 
symmetry sector, the number of such constraints is much smaller. 
Consider in particular the sector $(m, T) = (\tfrac12, 1)$.
This is a six dimensional Hilbert space and, as shown in the 
Appendix \ref{appb}, there are six symmetry-preserving constraints,
which can be written as a linear superposition of the 
independent occupation numbers:
\begin{align}
d^s_j(\nn) = \frac14 
\sum^2_{i = 0} \left(\kappa^j_{i\uparrow} n_{i\uparrow} +
 \kappa^j_{i\downarrow} n_{i\downarrow}\right),
\end{align}
where $\kappa_{i\sigma}^j$ are integers.
In addition, we have the simpler spin constraints 
$\sum_{i=0}^3 n_{i\uparrow} =  2$  and $\sum_{i=0}^3 n_{i\downarrow} = 1$.

A more elementary example is the traslationally 
invariant version of the Borland-Dennis setting 
(i.e., the Hilbert space 
$(N,M) = (3,6)$: $\mathcal{H} = {\rm span}
\{\ket{\vecform{\varphi}_{123}},
\ket{\vecform{\varphi}_{145}},
\ket{\vecform{\varphi}_{246}},
\ket{\vecform{\varphi}_{356}} \})
$ whose sym\-metry-preserving constraints can be easily computed.
Indeed,
$f_{123} = \tfrac12(n_1 + n_2 + n_3 - 1)$,
$f_{145} = \tfrac12(n_1 - n_2 + 1 - n_3)$,
and 
$f_{246} = \tfrac12(1-n_1 + n_2 - n_3)$.
One can recognize in $f_{356} = \tfrac12(1 -n_1 - n_2 + n_3)
= \tfrac12(2 -n_1 - n_2 - n_4)$ the famous Borland-Dennis
representability condition for three-fermions in a six
dimensional one-par\-ticle Hilbert space, safe a 
normalization constant. Therefore, in this case 
we can make use of the generalized Pauli principle by 
writing $f_{356} = \tfrac12 \mathcal{D}(\nn)$.
This is nothing more than
the constraint that we have saturated (i.e., 
$2 -n_1 - n_2 - n_4 = 0$) in Sec.~\eqref{sec3}.
The other fermionic constraints 
$d^s_{\alpha}({\vecform n}) \equiv f_\alpha(\vecform n)$ 
are just sophisticated versions of the (normal) Pauli 
exclusion principle $0 \leq n_i \leq 1$, plus the ordering
$n_{i} \geq n_{i+1}$. Notice in passing that in this example
all the constraints can be rewritten as
\begin{align}
f_{ijk} = \tfrac12(n_i + n_j + n_k-1),
\label{dist}
\end{align}
which shows that each one of these fermionic constraints 
measures the distance (normalized to 1)
to the opposite facet to the polytope's vertex 
$n_i = n_j = n_k = 1$, and $n_l = 0$ if $l\notin\{i,j,k\}$. 
For example, 
$f_{356}$ measures the distance to the non-elementary 
facet $2 = n_1 + n_2 + n_4$. A similar reasoning holds 
for the constraints of the former example of three 
$\tfrac12$-fermions in four lattice sides $(3,4)$. We believe that
this geometrical picture of the time evolution of the 
occupation numbers and the fermionic constraints 
indicates a promising future research path.

By means of the inversion of 
$\vecform{n} = \mathcal{M}^s_{(N,M)} \vecform{f}$,
we can now assign square amplitudes to 
constraints in a meaningful way (i.e, $f_\alpha = d^s_\alpha(\nn)$).
This last relation is exact at $t = 0$. Since the 
natural orbitals retain their orthonormality through the 
whole system's time evolution, it is also exact instantaneously.
Therefore, by employing the results of the last 
section, in particular Eq.~\eqref{eq:tdNON1}, 
the time-evolution of the natural occupation numbers is thus given by:
\begin{align}
\label{main}
\dot{\nn} = \mathcal{M}^s_{(N,M)} \dot{\vecform{f}},
\end{align}
where 
\begin{align}
\label{main2}
i\dot{f}_\alpha = \sum_{\beta \neq \alpha} \sqrt{d^s_\alpha(\nn)d^s_\beta(\nn)}
\left[
\bra{\tilde{\vecform{\varphi}}_\beta} \hat H \ket{\tilde{\vecform{\varphi}}_\alpha}
e^{i(\xi^{\rm dyn}_\beta - \xi^{\rm dyn}_\alpha)} - {\rm c.c.}\right].
\end{align} 
 The dynamical phase $\xi^{\rm dyn}_\alpha(t)$ is given 
by the Eq.~\eqref{coorph}, making again the 
substitution $f_\alpha = d^s_\alpha(\nn)$:
\begin{align}
\xi^{\rm dyn}_\alpha(t) = \int_0^t \sum_{\beta} 
\frac12 \sqrt{\frac{d^s_\beta(\nn)}
{d^s_\alpha(\nn)}} \left[
\bra{\tilde{\vecform{\varphi}}_\alpha} \hat H \ket{\tilde{\vecform{\varphi}}_\beta}
e^{i(\xi^{\rm dyn}_\alpha - \xi^{\rm dyn}_\beta)} + {\rm c.c.} \right] d\tau.
\label{main3}
\end{align}
In the case of the Hubbard model with three spin-${\tfrac12}$ 
fermions on four lattice sites, the evolving occupation 
numbers $n_{l\sigma}(t)$ lost the connection
with the orbital $\tilde\varphi_{l\sigma}$ as soon as the 
perturbation is switched on. Yet $n_{l\sigma}(t)$ is defined 
as the occupation number associated with the 
time evolving natural orbital $\tilde\varphi_{l\sigma}(t)$, such that 
$\tilde\varphi_{l\sigma}(0^+) =  \tilde\varphi_{l\sigma}$ and 
$n_{l\sigma}(0^+) = n_{l\sigma}$.

These three last equations are the main results of this paper. 
The time-evolution picture is then as follows. 
From the BBGKY hierarchy one can obtain Eq.~\eqref{eq:NONt} 
for the time evolution of the (phase-shifted) natural orbitals. 
From the time-dependent quantum mechanical action we obtained 
the equation of motion of the square amplitudes~\eqref{main2} 
and the relative phases~\eqref{main3}.
By exploiting generalized Pauli constraints 
(in a given symmetry sector) we have found Eq.~\eqref{main}
and its corresponding inversion 
for the time evolution of the natural occupation numbers. 
All the equations are written in terms of one-particle 
quantities (i.e., occupation numbers and natural orbitals)
plus a set of supplementary dynamical
phases (as many as independent occupation numbers),
which retain the memory effects of the system.

Four key observations are in order here. First, the 
phases $\vecform{\xi}$ are not attached to the 
natural orbitals in the sense of the so-called 
phase-included natural orbital theory \cite{Klaas}, 
and therefore our expressions go well beyond the 
realm of two-electron systems.  Second, 
Eqs.~\eqref{main} and \eqref{main2} can be understood as the
time-evolution of the recently discovered 
density-matrix functional for translationally 
invariant systems with periodic boundary conditions 
\cite{PhysRevLett.122.013001}. The divergence 
($1/\sqrt{D_\alpha(\nn)}$) observed in such a functional 
at the level of ground states is also observed here at 
the level of the dynamical phase for time evolving systems. 
Third, it is expected that Eqs.~\eqref{main} and \eqref{main2} 
are a reasonable approximation for the time evolution of 
molecular systems, because the contribution of single (and more 
generally odd) excitations in the configuration-interaction 
expansion in the natural-orbital basis tends to be 
negligible~\cite{Mentel2014, BenavQuasi2}. Finally, it is always
possible that even after recognizing of the symmetries satisfied 
by the ground (or initial) state the number of Slater determinants  
exceeds the number of natural orbitals. In that case one can 
always use the structural simplification due to pinning to 
reduced the dimensionality of the Hilbert space,
or to resort to more involve ``inversion'' techniques \cite{PhysRevA.97.052105}.

\section{Conclusions}

Describing the dynamics of strongly driven electron dynamics is a 
challenging problem. Multiconfigurational wave functions with enough 
terms to adequately describe correlation have so far been 
limited to small systems or short simulation times. 
In this paper we investigated the time evolution 
of the one-body reduced density matrix, based on recent 
progress on fermionic exchange symmetry for pure systems. 
In particular, we have employed the stability (under any small perturbation 
of the Hamiltonian) of the structural 
simplification of the wave function due to quasipinning 
or, more generally, symmetries. 
We presented two important results. First, we 
developed a closed expression for the time evolution of fermionic 
natural occupation. This evolution depends on one-particle quan\-tities 
(the natural orbitals and the occupation numbers themselves)
as well as to a set of dynamical phases. 
Second, we have presented a formula for the evolution of
such dynamical phases. We believe that this last equation is the missing
piece for the successful application of reduced density-matrix
functional theory to time-dependent problems.  
Since our approach alleviates the computational burden of the many-body
problem, we think that it can be an important tool to 
understand time-evolving strongly-correlated fermionic systems, 
whose physics is receiving increased attention \cite{Markus}. 

In this paper we have also given an estimate of the linear 
response of the distance to the boundaries of the polytope. In first 
order of the perturbation, quasipinning scales as
the ex\-pected value $\langle \hat D^2_j\rangle$,
which is zero only for pinned systems. In addition, 
since the symmetry-adapted fermionic constraints measures 
the distance (normalized to 1) to the polytope's 
facets \eqref{dist}, our work results in a remarkable 
geometrical picture
for the time evolution of highly correlated pure fermionic systems,
where fractional occupation are crucial for describing their dynamics. 
We think that this is a promising future research avenue.

We also think that our results further underline the important role 
played by symmetries in molecular theories of the one-body 
reduced density matrix
\cite{Maciek2017, DAVIDSON201328, doi:10.1063/1.4974096, PhysRevA.99.042516}.
All our findings can be rewritten in linear response theory, such 
that the computational burden can be further relieved. Work 
along these lines is already in progress.

\begin{acknowledgments}
We thank Nektarios Lathiotakis, Jamal Berakdar, 
Ryan Requist, and Christian Schilling for helpful 
discussions.

\end{acknowledgments}

\appendix 

\section{Linear response of quasipinning}
\label{appaa}

Let us consider a Hamiltonian $\hat H$ with 
non-degenerated eigenstates $\ket{\Psi_n}$ 
and eigenenergies $E_0 < E_1 \leq \cdots$. Let us write the eigenstates
in a complete basis of Slater determinants of 
(non-degenerated) ground-state
natural orbitals, such that:
\begin{align}
\label{app:sd}
\ket{\Psi_n} = \sum c^n_{\vecform{\alpha}}\ket{\vecform{\varphi_\alpha}},
\end{align}
where $\alpha$ stands for a string of numbers indexing the 
(non-degenerated) ground-state natural orbitals. Let us call
$\nn_0$ the vector of natural occupation numbers of the 
ground state $\ket{\Psi_0}$. For a given generalized Pauli 
constraint $\mathcal{D}_j(\nn_0)$ there is the operator 
\begin{align}
\hat D_j = \mathcal{D}_j(\hat n_{\varphi_1},\dots,\hat n_{\varphi_m}), 
\label{kappa}
\end{align}
where $\hat n_{\varphi_m}$ is the particle number operator for the ground-state
natural orbital $\ket{\varphi_j}$. By construction, every Slater determinant is 
an eigenvector of $\hat D_j$, namely,
\begin{align}
\label{ww}
\hat D_j \ket{\vecform{\varphi_\alpha}} = \kappa^j_\alpha \ket{\vecform{\varphi_\alpha}},
\end{align}
where $\kappa^j_\alpha$ is an integer.

In perturbation theory the ground state of the perturbed Hamiltonian 
$\hat H(\lambda) \equiv \hat H 
+ \lambda \hat V$ reads as $\ket{\Psi^{\lambda}_0} = \ket{\Psi_0} 
 - \lambda \sum_{n > 0} b_n 
\ket{\Psi_n} + \mathcal{O}(\lambda^2)$, with 
$b_n = \bra{\Psi_n}\hat V\ket{\Psi_0}/(E_n - E_0)$. We have
\begin{align}
\hat D_j \ket{\Psi^{\lambda}_0} = \hat D_j \ket{\Psi_0}
 - \lambda \sum_{n> 0} b_n 
\hat D_j \ket{\Psi_n} + \mathcal{O}(\lambda^2).
\end{align}
The evolution of the distance to the chosen polytope's facet is 
$\mathcal{D}_j(\nn_\lambda) = \bra{\Psi^{\lambda}_0}
\hat D_j \ket{\Psi^{\lambda}_0}$. Therefore
\begin{align}
\label{eq:appa}
\mathcal{D}_j(\nn_\lambda) = \mathcal{D}_j(\nn_0)
 - \lambda \sum_{n > 0} \left(b_n 
\bra{\Psi_0} \hat D_j \ket{\Psi_n} + c.c.\right) 
+  \mathcal{O}(\lambda^2).
\end{align}
Making use of Eq.~\eqref{app:sd} and \eqref{ww}, and the 
orthonormality of the Slater determinants, the second term 
in the r.h.s.~of \eqref{eq:appa} can be written as
\begin{align}
\label{eq:appb}
\sum_{n > 0, \alpha} b_n 
\bra{\Psi_0} \hat D_j \ket{\Psi_n} 
 = \sum_{n> 0, \alpha} b_n 
(c^{0}_\alpha)^* c^n_\alpha \kappa^j_\alpha.
\end{align}
Taking the square of the absolute value of the r.h.s.~of \eqref{eq:appb}
\begin{align}
\label{eq:appabs}
\left|\sum_{n> 0, \alpha} b_n 
(c^{0}_\alpha)^* c^n_\alpha \kappa^j_\alpha\right|^2
\leq 
\left(\sum_{\alpha} |\gamma_\alpha|^2  \right)
\left(\sum_{\alpha} 
|c^{0}_\alpha|^2 (\kappa^j_\alpha)^2\right) ,
\end{align}
where $\gamma_\alpha = \sum_{n> 0} b_n c_\alpha^n$ is a relative 
strength. 
In the last inequality we have employed the Cauchy-Schwarz
inequality for the ``vectors'' $[u]_\alpha \equiv \kappa^j_\alpha c^0_\alpha$ and 
 $[v]_\alpha \equiv \sum_{n>0} b_n c^n_\alpha$. 

Finally, by noticing that $|u|^2$ is the expectation
value of the square of $\hat D_j$, namely,
$\sum_{\alpha} 
|c^{0}_\alpha|^2 (\kappa^j_\alpha)^2 = \bra{\Psi_0} \hat D^2_j \ket{\Psi_0} 
\equiv \langle\hat D^2_j \rangle$ we have the 
estimate 
\begin{align}
\mathcal{D}_j(\nn_\lambda) \leq \mathcal{D}_j(\nn_0)
 + 2 \lambda \gamma \sqrt{\langle\hat D^2_j \rangle} + \mathcal{O}(\lambda^2),
\end{align}
where $\gamma = \sqrt{\sum_{\alpha}|\gamma_\alpha|^2}$. Remarkably,
$\gamma$ can also be estimated as follows:
\begin{align}
\sum_{\alpha}|\gamma_\alpha|^2
&= \sum_\alpha \left(\sum_{n> 0} b_n c_\alpha^n\right)
\left(\sum_{n> 0} b_n c_\alpha^n\right)^* \nonumber \\
&= \sum_\alpha \sum_{n, m> 0} b_n b_m^* 
\bra{\Psi_m}\vecform{\varphi_\alpha}\rangle
\bra{\vecform{\varphi_\alpha}}\Psi_n\rangle = \sum_{n > 0} |b_n|^2.  
\nonumber
\end{align}
In the last line we have used the resolution of the identity in the 
basis of Slater determinants and the orthonormality of the eigenstates 
$\ket{\Psi_m}$. 

Finally, by noticing that 
\begin{align}
\sum_{n > 0} |b_n|^2  &= 
\sum_{n > 0} \frac{\bra{\Psi_0}\hat V \ket{\Psi_n}\bra{\Psi_n}\hat V \ket{\Psi_0}}{(E_n - E_0)^2}
\nonumber \\
&\leq 
\sum_{n > 0} \frac{\bra{\Psi_0}\hat V \ket{\Psi_n}\bra{\Psi_n}\hat V \ket{\Psi_0}}{(E_1 - E_0)^2}
\nonumber \\
&= \frac{\bra{\Psi_0}\hat V (1 - \ket{\Psi_0}\bra{\Psi_0})\hat V \ket{\Psi_0}}{(E_1 - E_0)^2} \nonumber\\
&= \frac{\bra{\Psi_0}\hat V^2 \ket{\Psi_0}
- \bra{\Psi_0}\hat V \ket{\Psi_0}^2}{(E_1 - E_0)^2} ,
\end{align}
we obtain 
\begin{align}
\gamma \leq \frac{\sqrt{{\rm Cov}_{\Psi_0}(\hat V)}}{E_{\rm gap}} ,
\end{align}
where the energy gap is defined as $E_{\rm gap} = E_1 - E_0$
and the co\-va\-riance ${\rm Cov}_{\Psi_0}(\hat V) =
\bra{\Psi_0}\hat V^2 \ket{\Psi_0}
- \bra{\Psi_0}\hat V \ket{\Psi_0}^2$.

\section{
Removing of the Slater-geometrical phase $\xi^{\rm geo}_\alpha$}
\label{appc}

Notice that one can expand the time-derivative of a natural orbital 
with respect to the complete set of natural orbitals:
\begin{align}
i \frac{d}{dt} \ket{\varphi_k} = \eta_{kk}(t) \ket{\varphi_k} + 
\sum_{m\neq k} \eta_{mk}(t) \ket{\varphi_m}.
\end{align}
Due to the orthonormality of the natural orbitals, 
$\eta_{mk}(t) \equiv \bra{\varphi_m}i \partial_t \ket{\varphi_k}$ is a Hermitian matrix.
Extracting the phase factor 
$\exp(-i\int_0^t \eta_{kk}(\tau) d\tau)$ from the natural orbital $\ket{\varphi_k}$ 
removes the diagonal terms $\eta_{kk}(t)$ such that now the time derivative 
of $\ket{\varphi_k}$ is completely determined by the non-diagonal 
elements of the matrix $\eta_{mk}(t)$. Indeed, by defining the 
phase-shifted natural orbitals
$\ket{\tilde{\varphi}_k} = \exp(i\int_0^t \eta_{kk}(\tau) d\tau) \ket{\varphi_k}$
(and consequently $\tilde \eta_{mk} = \bra{\tilde \varphi_m}i \partial_t \ket{\tilde \varphi_k}$),
one obtains $\bra{\tilde\varphi_k}\partial_t\ket{\tilde\varphi_k} = 0$. 
For Slater determinants:
\begin{align}
\ket{\vecform{\varphi}_{\alpha}} &\equiv\ket{\varphi_{\alpha_1}\dots\phi_{\alpha_m}} \nonumber \\
 &= e^{-\sum_j i\int_0^t \eta_{\alpha_j\alpha_j} d\tau} \nonumber
\ket{\tilde\varphi_{\alpha_1}\dots\tilde\phi_{\alpha_m}} 
\\ &=  \nonumber e^{-i \int_0^t
\bra{\vecform{\varphi}_{\alpha}}i\partial_{\tau}\ket{\vecform{\varphi}_{\alpha}}d\tau} \ket{\vecform{\tilde\varphi}_{\alpha}}
 \\ &= e^{i\xi^{\rm geo}_\alpha} \ket{\vecform{\tilde\varphi}_{\alpha}}.
\end{align}
Finally, it is an elementary exercise to verify that, as a consequence of 
the orthonormality of the natural orbitals, the phase-shifted natural 
orbitals are also orthonormal along the whole time evolution of the system
\begin{align}
\bra{\tilde\varphi_k(t)}\tilde\varphi_j(t)\rangle = \delta_k^j.
\end{align}

\section{Hubbard model $(N = 3, L = 4)$ }
\label{appb}

Consider three spin-$\tfrac12$ fermions
on four lattice  sites. Consider also
the symmetry sector $(m, T) = (\tfrac12, 1)$.
To determine all Slater determinants
$\ket{\kappa_1 m_1, \kappa_2 m_2, \kappa_3 m_3}$ 
with total momentum $K = \tfrac{2\pi}4 \sum_i \kappa_i ({\rm mod} 4)$ 
and total magnetization $M_z = \sum_n m_n = \tfrac12$ is straightforward.
Using $\uparrow, \downarrow$ for the spin coordinates one obtains
six states: 
\begin{align*}
\ket{\vecform{\varphi}_{001}} &=
\ket{0\uparrow 0 \downarrow 1 \uparrow}, \\
\ket{\vecform{\varphi}_{113}} &=
\ket{1\uparrow 1 \downarrow 3 \uparrow}, \\
\ket{\vecform{\varphi}_{221}} &=
\ket{2\uparrow 2 \downarrow 1 \uparrow}, \\
\ket{\vecform{\varphi}'_{023}} &= 
\ket{0\uparrow 2 \downarrow 3 \uparrow} \\
\ket{\vecform{\varphi}''_{023}} &=
\ket{0\uparrow 2 \uparrow 3 \downarrow}, \\
\ket{\vecform{\varphi}'''_{023}} &=
\ket{0\downarrow 2 \uparrow 3 \uparrow}.
\end{align*}
Since occupation numbers satisfy two constraints, namely,
$n_{0\uparrow} + n_{1\uparrow} + n_{2\uparrow} + n_{3\uparrow} = 2$
and 
$n_{1\downarrow} + n_{2\downarrow} + n_{2\downarrow} + n_{3\downarrow} = 1$,
there are only six independent occupation numbers which can be related with 
the corresponding square amplitudes:
\begin{align*}
\begin{pmatrix}
n_{0\uparrow} \\
n_{0\downarrow} \\
n_{1\uparrow} \\
n_{1\downarrow} \\
n_{2\uparrow} \\
n_{2\downarrow} 
\end{pmatrix} 
= 
\begin{pmatrix}
1 & 0 & 0 & 1 & 1 & 0 \\
1 & 0 & 0 & 0 & 0 & 1 \\
1 & 1 & 1 & 0 & 0 & 0 \\
0 & 1 & 0 & 0 & 0 & 0 \\
0 & 0 & 1 & 0 & 1 & 1 \\
0 & 0 & 1 & 1 & 0 & 0 
\end{pmatrix} 
\begin{pmatrix}
f_{001} \\
f_{113} \\
f_{221} \\
f'_{023} \\
f''_{023} \\
f'''_{023} 
\end{pmatrix}.
\end{align*}
By inverting the matrix one obtains six symmetry-preserving 
generalized Pauli constraints, namely:
\begin{align*}
d_1^s(\nn) &= \tfrac14(n_{0\uparrow}+ n_{0\downarrow} + 2 n_{1\uparrow}
-2 n_{1\downarrow} - n_{2\uparrow} - n_{2\downarrow}), \\
d_2^s(\nn) &= n_{1\downarrow}, \\
d_3^s(\nn) &= \tfrac14(-n_{0\uparrow} - n_{0\downarrow} + 2 n_{1\uparrow}
-2 n_{1\downarrow} + n_{2\uparrow} + n_{2\downarrow}), \\
d_4^s(\nn) &= \tfrac14(n_{0\uparrow}+ n_{0\downarrow} - 2 n_{1\uparrow}
+ 2 n_{1\downarrow} - n_{2\uparrow} +3 n_{2\downarrow}), \\
d_5^s(\nn) &= \tfrac14(2n_{0\uparrow} -2 n_{0\downarrow}  
+2 n_{2\uparrow} - 2n_{2\downarrow}), \\
d_6^s(\nn) &= \tfrac14(-n_{0\uparrow}+ 3n_{0\downarrow} - 2 n_{1\uparrow}
+2 n_{1\downarrow} + n_{2\uparrow} + n_{2\downarrow}). 
\end{align*}
Notice the normalization factor $\tfrac14$ appearing in front of the 
constraints (hence, $0 \leq d^s_j(\nn) \leq 1$). Obviously, the square 
amplitudes and the symmetry-preserving constraints are related:
$$
f_{001} = d^s_1(\nn), \quad f_{113} = d_2^s(\nn),
\quad f_{221} = d_3^s(\nn)
$$ 
and so on. Finally, to make things easier
we are not imposing additional symmetries, like the total spin operator.

\bibliography{Time}

\end{document}